# Motion Compensated Whole-Heart Coronary Magnetic Resonance Angiography using Focused Navigation (fNAV)

Christopher W Roy[1], John Heerfordt[1,2], Davide Piccini[1,2], Giulia Rossi[1], Anna Giulia Pavon[3], Juerg Schwitter[3,4,5], Matthias Stuber[1,6]

[1]Department of Radiology, Lausanne University Hospital (CHUV) and University of Lausanne (UNIL), Switzerland
[2]Advanced Clinical Imaging Technology (ACIT), Siemens Healthcare AG, Lausanne, Switzerland
[3]Division of Cardiology Lausanne University Hospital (CHUV), Lausanne, Switzerland
[4]Director CMR-Center Lausanne University Hospital (CHUV), Lausanne, Switzerland
[5]Faculty of Biology and Medicine, University of Lausanne (UNIL), Lausanne, Switzerland
[6]Center for Biomedical Imaging (CIBM), Lausanne, Switzerland

**Submitted to Journal of Cardiovascular Magnetic Resonance**

**Grant support: Schweizerischer Nationalfonds zur Förderung der Wissenschaftlichen Forschung, Grant/Award Number: #173129.**

**Corresponding Author:**
   Christopher W. Roy, PhD
   0000-0002-3111-8840
   Lausanne University Hospital (CHUV)
   Rue de Bugnon 46, BH-7-84
   1011 Lausanne, Switzerland
   Phone: +41 21 314 7516
   Fax: +41 21 314 4443
   christopher.roy@chuv.ch

# Abstract

### Background

Radial self-navigated (RSN) whole-heart CMRA is a free-breathing technique that estimates and corrects for respiratory motion. However, RSN has been limited to a 1D rigid correction which is often insufficient for patients with complex respiratory patterns. The goal of this work is therefore to improve the robustness and quality of 3D radial CMRA by incorporating both 3D motion information and nonrigid intra-acquisition correction of the data into a framework called focused navigation (fNAV).

### Methods

We applied fNAV to 500 data sets from a numerical simulation, 22 healthy volunteers, and 549 cardiac patients. In each of these cohorts we compared fNAV to RSN and respiratory resolved XD-GRASP reconstructions of the same data. Reconstruction times for each method were recorded. Motion estimate accuracy was measured as the correlation between fNAV and ground truth for simulations, and fNAV and image registration for *in vivo* data. Vessel sharpness was measured using Soap-Bubble. Finally, subjective image quality analysis was performed by a blinded expert reviewer who chose the best image for each *in vivo* data set.

### Results

The reconstruction time for fNAV images was significantly higher than RSN (6.1 ± 2.1 minutes vs 1.4 ± 0.3, minutes, p<0.025) but significantly lower than XD-GRASP (25.6 ± 7.1, minutes, p<0.025). Overall, there is high correlation between the fNAV, and reference displacement estimates across all data sets (0.73 ± 0.29). For simulated data, volunteers, and patients, fNAV lead to significantly sharper coronary arteries than all other reconstruction methods (p < 0.01). Finally, in a blinded evaluation by an expert reviewer fNAV was chosen as the best image in 239 out of 571 data sets (p = $10^{-5}$).

## Conclusion

fNAV is a promising technique for improving the quality of self-navigated free-breathing 3D radial whole-heart CMRA. This novel approach to respiratory self-navigation can derive 3D nonrigid motion estimations from an acquired 1D signal yielding statistically significant improvement in image sharpness relative to 1D translational correction as well as XD-GRASP reconstructions. Further study of the diagnostic impact of this technique is therefore warranted to evaluate its full clinical utility.

## Background

Whole-heart Coronary Magnetic Resonance Angiography (CMRA) is a non-invasive alternative to X-ray coronary angiography, providing high resolution assessment of complex cardiac structures without exposure to ionizing radiation (1,2). In conventional CMRA, electrocardiogram triggering is used to limit data collection to mid-diastole, effectively "freezing" cardiac motion (3). Additionally, a one dimensional (1D) navigator echo is typically prescribed over the dome of the liver to monitor the primary direction of respiratory motion (4,5) and further limit data collection to a small acceptance window manually defined at end-expiration. Unfortunately, prospective gating is often impeded by physiological variability, resulting in inefficient and unpredictable scan times which has led to several proposed strategies for free-breathing CMRA.

Among the alternatives to conventional prospective respiratory motion tracking and data rejection is radial self-navigated (RSN) whole-heart CMRA (6). In RSN CMRA, three-dimensional (3D) radial k-space data are acquired throughout the entire respiratory cycle and a readout orientated along the superior-inferior (SI) direction is repeated at the beginning of each radial interleave. In this way, the movement of the heart due to respiration can be quantified and corrected for in a patient-specific manner by measuring the relative correlation between Fourier transformed SI readouts over a region of interest containing the blood pool. RSN CMRA, using spiral phyllotaxis distribution of the radial readouts (7,8), has been shown to provide high-quality images with isotropic resolution in patient studies (9–

11) and improved scanning efficiency relative to conventional prospectively navigated Cartesian CMRA (12).

Still, the existing approach for RSN is limited to respiratory motion quantification in one-dimension (1D) along the SI direction which is often insufficient for patients with significant respiratory motion along the anterior-posterior or left-right directions. Furthermore, subsequent rigid correction of the k-space data can not account for the non-linear behavior of respiratory motion. As a result, an alternative reconstruction of 3D radial CMRA data was proposed wherein a respiratory signal is derived from repeated SI readouts, the k-space data are sorted in to multiple respiratory states and reconstructed as respiratory resolved images using eXtra-Dimensional Golden-angle RAdial Sparse Parallel MRI (XD-GRASP) (13,14). While it has been shown that XD-GRASP provides sharper images than the 1D correction scheme, this approach may be adversely affected by residual uncorrected intra-bin motion, overregularization, and long computation times (15).

The goal of this work is therefore to improve the robustness and quality of respiratory self-navigated 3D radial CMRA by incorporating both 3D motion information and nonrigid intra-acquisition correction of the data. To do this, we propose a novel framework hereafter referred to as focused navigation (fNAV). Our approach integrates the following three features: (i) a 3D radial CMRA acquisition with periodically repeated SI readouts, (ii) an autofocusing-based algorithm (16–20) that converts a unitless 1D respiratory signal derived from SI readouts into displacement fields along all three spatial dimensions with physical units, and (iii) an iterative reconstruction that optimizes both local image sharpness and smoothness in the displacement fields, resulting in a final 3D image that is regionally corrected for intra-acquisition respiratory motion.

To evaluate the fNAV framework, we present a comprehensive numerical simulation that provides ground truth references for displacement field estimates and image reconstructions. We then demonstrate the use of fNAV in healthy volunteers and in a large cohort of patients. Finally, we test the hypothesis that 3D nonrigid respiratory motion correction using fNAV improves coronary vessel sharpness relative to previously reported methods for 1D corrected RSN (8) and respiratory resolved XD-GRASP reconstructions of the same data sets (13,14).

## Methods

### Respiratory Signal Extraction

The fNAV framework is applied to a previously described prototype respiratory self-navigated ECG-triggered 3D balanced stead state free precession (bSSFP) sequence with spiral phyllotaxis radial sampling, T2-preperation, spatial pre-saturation, and fat saturation pulses (7–9). At the beginning of each radial interleave, a readout orientated along the SI direction is used for self-navigation (Fig. 1a). To extract a respiratory signal (Fig. 1b), principal component analysis is applied to a matrix of SI projections (Fourier transform of each SI readout) from each receiver coil. The strongest principal component within the expected respiratory frequency range is chosen as the respiratory signal (14,21).

### Focused Navigation

It is well established that SI-derived respiratory signals (6,8) are proportional to bulk motion within the sensitivity range of the receiver coils or the imaging volume for non-selective and selective excitations respectively. However, such signals are unitless and therefore require a patient-specific calibration or iterative algorithm to measure physical displacement (8,22). The proposed fNAV framework (Fig. 1c) posits that patient-specific 3D displacement can be approximated by a normalized SI-derived respiratory signal $S(t)$ measured at time points (t) multiplied with fNAV coefficients $\mathbf{A(r)} = [A_x(r), A_y(r), A_z(r)]$ that describe the millimetric amplitude of respiratory motion for a given spatial location $\mathbf{r} = [x, y, z]$. Using, these coefficients, an image ($I_r$) can be reconstructed with translational motion correction as follows:

$$I_r = FK(t)e^{i2\pi \mathbf{k(t)} \cdot \mathbf{A(r)} \cdot S(t)} \qquad (1)$$

where (K) is the acquired 3D radial k-space data with coordinates $\mathbf{k(t)} = [k_x(t), k_y(t), k_z(t)]$, and (F) is the non-uniform Fourier transform. It is important to note that by correcting k-space with Eq. 1 the resulting image will be sharper in regions where $\mathbf{A(r)}$ reflects the underlying motion state (i.e. heart moving due to respiration) but become blurrier where it does not (i.e. static tissue). To address this, we reconstruct fNAV images (I) using localized linear translations (16):

$$I = \sum_r U_r I_r. \qquad (2)$$

For each spatial location, a motion corrected image $I_r$ is reconstructed using Eq. 1 and the corresponding intensity value is added to the fNAV image using a mask ($U_r$) which contains zeros everywhere except for the selected spatial location. In this way, every voxel in the resulting fNAV image is regionally corrected for respiratory motion. Similarly, we reconstruct fNAV displacement fields (**D**) as follows:

$$\mathbf{D} = \sum_r U_r \mathbf{A}(\mathbf{r}) \qquad (3)$$

where each spatial location in D contains the corresponding fNAV coefficient.

To implement Eqs. 1-3, reconstruct fNAV images, and reconstruct fNAV displacement fields, we must estimate fNAV coefficients for every sampled spatial location. To do this in a computationally efficient manner, we propose a three-step approach. First, we estimate fNAV coefficients **A(r̃)** where r̃ specifies a small region-of-interest containing the heart. Second, we use our estimate of **A(r̃)** to constrain solutions for the remaining fNAV coefficients. Third, we iteratively refine all of our fNAV coefficients by enforcing smooth transitions in the corresponding fNAV displacement fields.

**Estimating displacement of the heart**

Beginning with an initial estimate **A(r̃) = [0,0,0]**, an intermediate image (Ĩ) is reconstructed using Eq. 1, and an image quality metric is used to iteratively improve the estimate for **A(r̃)** (Fig. 1c). This approach is well-known as autofocusing (23–26). For the fNAV framework we use the previously validated metric of localized image gradient entropy H:

$$H = -\sum_{u=x-\frac{b}{2}}^{x+\frac{b}{2}} \sum_{v=y-\frac{b}{2}}^{y+\frac{b}{2}} \sum_{w=z-\frac{b}{2}}^{z+\frac{b}{2}} p_{uvw} \log_2(p_{uvw}) \qquad (4)$$

$$p_{uvw} = \frac{g_{uvw}}{\sum_{uvw} g_{uvw}}$$

$$g_{uvw} = \sqrt{|\nabla_u \tilde{I}|^2 + |\nabla_v \tilde{I}|^2 + |\nabla_w \tilde{I}|^2}$$

where (b) defines the main lobe width of a separable low pass Hanning filter centered around a given spatial location, (p) is the normalized voxel intensity from the gradient (g) of the intermediate image, and $\nabla$ is approximated by 1D finite differences (16,17,19,26). The value of **A(r̃)** that minimizes H is solved using a steepest descent algorithm where the gradient of H as a function of **A(r̃)** is approximated numerically:

### *Estimating regional displacement*

The second step of the fNAV framework is to estimate fNAV coefficients for the remaining spatial locations**.** In principle we could perform the same iterative optimization described by the previous section and Fig. 1c, for every region of our 3D image. However, to reduce the computational burden, we instead form a bank ($B_A$) of fNAV coefficients, image reconstructions ($B_I$), and metric values ($B_H$), corresponding to all of the tested values from the previous step, as well as a small grid of values near the optimum coefficients for a total of (m) motion states (Fig. 1d). Now, in place of the steepest descent algorithm, the value of **A(r)** that minimizes H is efficiently solved by a sorting algorithm applied to $B_H$ and $B_A$.

### *Refining motion estimates*

The third and final step of the proposed framework is a joint optimization of the image metric and a smoothness constraint applied to the fNAV displacement fields (**D**) calculated from Eq 4:

$$\mathbf{A(r)} = \arg\min_m (H + \lambda \, |\mathbf{C(r)} - \mathbf{A(r)}|) \tag{5}$$

$$\mathbf{C(r)} = \frac{1}{c^3} \sum_{u=x-\frac{c}{2}}^{x+\frac{c}{2}} \sum_{v=y-\frac{c}{2}}^{y+\frac{c}{2}} \sum_{w=z-\frac{c}{2}}^{z+\frac{c}{2}} \mathbf{D}(uvw) \tag{6}$$

where **C(r)** defines an average operator over a sliding window (c), applied to **D**. The metric term (H) is first solved using $B_H$ and $B_A$ as previously described, and **D** derived from that solution is iteratively updated using eq. 4 (Fig. 1e). Finally, the refined estimates of **A(r)** are used to create a final 3D image with corrected respiratory motion (Fig. 1f).

*Tuning parameters*

The fNAV framework contains four user-defined "tuning" parameters: (i) the number of motion states (m) added to $B_A$, $B_I$ and $B_H$, (ii) the width (b) of the localized image metric defined in eq. 4, (iii) the width (c) of the smoothness constraint for the displacement fields defined in eq. 6, and (iv) the weighting parameter λ defined in eq. 5 which provides a trade-off between the image metric and variation in the fNAV displacement fields. To investigate the impact of these parameters and determine their optimum values a comprehensive numerical simulation was developed and used for a non-exhaustive search focused on a trade-off between computation time and accuracy in the respiratory motion estimations.

## Numerical Simulation Framework

Self-navigated ECG-triggered 3D radial data are synthesized using a numerical simulation developed for this work and inspired by the previously described MRXCAT approach (27). In summary, high resolution (1 mm$^3$) 3D volumes covering the chest are derived from the XCAT software which contains labels for each tissue of interest and produces realistic nonrigid cardiac and respiratory motion (28). For a user-defined maximum level of respiratory motion, a total of 400 unique volumes are generated from XCAT and arranged into a five-dimensional (5D) array representing the 3D volume sampled across 20 phases of a full cardiac and respiratory cycle. The ground truth respiratory motion-fields are also generated from XCAT.

To create synthetic physiological data, cardiac cycles with realistic heart-rate variability and respiratory cycles with variability in both frequency and amplitude are generated, spanning the length of a synthetic MR acquisition (29). The user-defined acquisition parameters for the synthetic MR sequence (Table 1) are chosen to match the *in vivo* data acquisitions described in the following section. For a given timepoint in the synthetic acquisition, a 3D volume representing the desired cardiac and respiratory phase is interpolated from the 5D array described above, the labelled tissues are converted to MR contrast using relaxation properties from the literature and a bSSFP signal equation, and the inverse NUFFT, which contains simulated 3D coil sensitivities, is used to extract the desired radial readout (29). Finally, complex gaussian noise is added to the synthetic k-space data.

## Numerical Simulation Data Acquisition

To optimize the tuning parameters of the fNAV algorithm, validate displacement field estimations, and validate image reconstructions, 500 synthetic data sets were generated using the framework described above. The maximum respiratory motion amplitude ranged from 0-5 mm, 0-10 mm, and 5-20 mm along the x (LR), y (AP), and z (SI) directions, respectively.

## Healthy Volunteer and Patient Data Acquisition

In this retrospective study, *in vivo* data were analyzed from 22 healthy volunteers (7 female, age 27 ± 5) and 549 cardiac patients (196 female, average age 58 ± 18) that were scanned at our institution between May 2014 and May 2016 on a 1.5T clinical MRI scanner (MAGNETOM Aera, Siemens Healthcare, Erlangen, Germany) with the previously described prototype sequence and scan parameters listed in Table 1 (9). All participants provided written informed consent in accordance with our institutional guidelines.

## Image Reconstruction

To study the performance of fNAV, all the simulated plus the 571 in vivo human data sets were reconstructed using 4 different approaches. In addition to the fNAV reconstruction described above and in Fig. 1, images were reconstructed without motion correction, using a previously reported method for 1D correction of self-navigation data (8), and using respiratory resolved XD-GRASP (14). The XD-GRASP reconstruction matched previously reported parameters with total variation along the respiratory dimension (weight: 0.05), and solved using a conjugate gradient algorithm with 20 iterations (13,14). All reconstructions and analyses were performed in MATLAB (The MathWorks, Inc., Natick, MA, USA) on a workstation equipped with two Intel Xeon CPUs, 512GB of RAM, and an NVIDIA Tesla GPU. To increase the computation speed for fNAV reconstruction parallel processing was used to reconstruct individual volumes for the image bank. For all *in vivo* data sets, reconstruction times were recorded for RSN, XD-GRASP, and fNAV reconstructions and compared using paired t-tests with the Bonferroni correction for multiple comparisons

## Motion Estimation Analysis

To assess the accuracy of fNAV estimates of 3D nonrigid motion, the optimized fNAV motion maps were visually compared to the ground truth values for the simulated data sets, and to motion maps derived from co-registering the end-inspiratory and end-expiratory bins of the XD-GRASP reconstructions for volunteer and patient data using NiftyReg (30). The quality of the fNAV motion maps was then assessed by a linear fit and Pearson correlation coefficient for the x, y, and z components across four regions of interest approximately defined over the aortic arch, base of the heart, apex of the heart, and liver.

## Image Quality Analysis

To assess the impact of motion correction on both simulated and *in vivo* image reconstruction, quantitative assessment of image quality across all reconstructions was assessed by the percentage vessel sharpness and visible vessel length of the left main + anterior descending (LAD), left circumflex (LCX), and right (RCA) coronary arteries using Soap-Bubble (31). For simulated images, this process was automated using the known vessel locations and therefore all 500 simulated data were assessed for vessel sharpness but not for vessel length. Conversely, all 22 healthy volunteers were manually examined using Soap-Bubble as was a subset of 20 randomly selected patients. All measurements were statistically compared using paired t-tests with the Bonferroni correction for multiple comparisons.

For a given volunteer or patient, images corresponding to the four reconstruction methods were placed in random order and the best reconstruction method was identified by a blinded expert reviewer (DP) with ten years experience in CMRA. This analysis was performed on all 571 *in vivo* data sets and statistical significance was measured using a chi-square test. Finally, two patients who underwent coronary angiography were chosen to provide a qualitative comparison between x-ray angiography and fNAV images.

# Results

## Tuning Parameters

Using the numerical simulation, the optimum size of the coefficient, image, and, metric banks corresponding to the number of motion states (m) considered in the fNAV reconstruction, was determined by a linear distribution between the uncorrected state (i.e. $A_x = A_y = A_z = 0$) and the optimum bulk fNAV coefficient as described in Fig. 1. This set-up was empirically observed to provide adequate representation of the nonrigid components of the ground truth simulated respiratory motion without the need for exhaustive searches. Additionally, the width of the localized image metric (b = 30 mm), width of the smoothness constraint (c = 7 mm), and weighting parameter ($\lambda$ = 0.75) provided the highest accuracy for respiratory motion estimation.

## Image Reconstruction

The reconstruction time for fNAV images (mean and standard deviation) was significantly higher than RSN (6.1 ± 2.1 minutes vs 1.4 ± 0.3, minutes, p<0.025) but significantly lower than XD-GRASP (6.1 ± 2.1 minutes vs 25.6 ± 7.1, minutes, p<0.025). Variability in reconstruction time was primarily due to the number of active coil elements, matrix size, total number of acquired lines, and number of motion states present in the final fNAV reconstruction.

## Motion Estimation Analysis

Fig. 2 compares representative fNAV displacement fields to reference displacement fields during end-inspiration. For simulated data (Fig. 2a-c), the reference is the ground truth, whereas for healthy volunteer (Fig. 2d-f) and patient data (Fig. 2g-i) the reference corresponds to displacement fields derived from registering XD-GRASP images. Overall, there is good visual agreement between fNAV displacement fields and the corresponding reference displacement fields.

For further evaluation of fNAV motion estimates, Fig. 3 plots the mean x, y, and z components of the fNAV and reference displacement fields in ROIs containing the aortic arch, base and apex of the heart, and liver, for each simulated (Fig.3a-d), healthy volunteer (Fig.3e-h), and patient (Fig.3i-l) data set. Overall, there is high correlation between the fNAV, and reference displacement estimates in particularly for the primary z direction in regions containing the heart. At the periphery of the image there is somewhat weaker correlation as well as for small values of displacement.

**Image Quality Analysis**

Fig. 4 shows curved reformatted images of the LAD (blue arrows) and RCA (red arrows) from reconstructions of simulated data across three levels of respiratory motion. At low levels of motion (Fig. 4a: maximum amplitudes LR = 0 mm, AP = 0 mm, SI = 10 mm), the self-navigated (second column), XD-GRASP (third column), and fNAV (fourth column) provide clear visual improvement over the uncorrected images (first column) and the resulting images are similar to the ground truth reference (fifth column). However, for increasing levels of respiratory motion (Fig. 4b: maximum amplitudes LR = 0 mm, AP = 5 mm, SI = 10 mm, Fig. 4c: maximum amplitudes LR = 5 mm, AP = 10 mm, SI = 20 mm), only fNAV and XD-GRASP provide comparable visual image quality to the reference with less noise but more blur in the XD-GRASP reconstructions due to regularization.

Fig. 5 provides a summary of vessel sharpness measurements across all 500 simulated data sets for the four reconstruction methods and three vessels. Overall, these simulated results corroborate previous studies showing that RSN provides a statistically significant increase in vessel sharpness relative to uncorrected images, that XD-GRASP images are significantly sharper than RSN, but we also see that fNAV leads to significantly sharper coronary arteries than all other reconstruction methods for the three vessels in keeping with the qualitative results shown in Fig. 4.

Fig. 6 shows curved reformatted images of the RCA (Fig. 6a), LAD (Fig. 6b), and LCX (Fig. 6c) from three representative healthy volunteers. Overall, these *in vivo* images corroborate the findings from the simulated data shown in Fig. 4 with fNAV providing the most consistent visualization of the full length of all three vessels. Similarly, Fig. 7 shows curved reformatted images of the RCA (Fig. 7a), and LAD+LCX (Fig. 7b-c), and LCX (Fig. 7c) from three representative patients. Once again, these images corroborate the findings from both the simulated data and healthy volunteers with fNAV providing the best conspicuity of all three observed coronary arteries.

Fig. 8 provides a summary of vessel sharpness and visible length measurements across all 22 healthy volunteer and a subset of 20 patient data sets for the four reconstruction methods and three vessels. Overall, fNAV provides the sharpest vessel measurements with statistically significant increases relative to all other reconstructions in both the first 4cm and full length of the RCA, LAD, and LCX ($p < 0.01$). Conversely, vessel length measurements were relatively consistent across the four reconstruction methods.

In the assessment of which reconstruction method was selected as the best image by an expert reviewer across all 571 *in vivo* data sets, RSN (n = 111) was chosen slightly more often than uncorrected (n = 99), XD-GRASP (n = 122) slightly more than RSN, but overall fNAV (n = 239) was observed to provide the best image quality in nearly half of the data sets ($p = 10^{-5}$).

Finally, Fig. 9 demonstrates the feasibility of using fNAV to visualize coronary artery disease through qualitative comparison to coronary angiography. In two patients, one with a significant stenosis (A-B) and one with a total occlusion (C-D) the disease is well visualized by the corresponding fNAV images.

## Discussion

In this work, we developed and validated a novel framework for nonrigid, regional intra-acquisition correction of respiratory motion in self-navigated 3D whole-heart CMRA scans. We showed that fNAV can accurately estimate and correct 3D respiratory motion yielding significant improvements in image quality and vessel sharpness relative to previously established approaches for 3D radial CMRA. Validation of the fNAV framework and tuning parameters were performed in a comprehensive numerical simulation. These numerical results were then corroborated by *in vivo* reconstructions of 22 healthy volunteer and 549 patient data sets demonstrating the robustness of our proposed reconstruction framework.

Our results agree with previous studies that used the same self-navigated sequence to demonstrate improved image quality when comparing respiratory-resolved XD-GRASP reconstructions to motion corrected reconstructions using a 1D translational model. Yet in our study, motion correction using fNAV consistently yielded the highest values for objective vessel sharpness measurement when compared to both RSN and XD-GRASP in simulations, healthy volunteers, and patients. Additionally, fNAV was selected as the best reconstruction with a statistically significant greater frequency in the more subjective yet blinded image quality assessment of the *in vivo* data. This suggests that not only is a 3D model important for respiratory motion correction but that within the range of simulated and observed *in vivo* motion, XD-GRASP may suffer from blur due to uncorrected intra-bin motion or over-regularization. Additionally, fNAV reconstruction could be performed significantly faster than XD-GRASP (6.1 ± 2.1 minutes vs 25.6 ± 7.1, minutes, $p<0.025$). In principle, the motion information derived from fNAV could be used to inform intra-bin correction of XD-GRASP reconstructions (32) or the fNAV displacement fields could be inserted directly into a compressed sensing or similar iterative denoising reconstruction (33). Such approaches may allow for a reduction in scan time albeit at the likely cost of increased computation time during image reconstruction.

Overall, fNAV builds on previous work using autofocusing to correct motion in MR images but this study is, to our knowledge, the first use of autofocusing for correcting free-breathing 3D radial self-navigated CMRA data. Cheng *et al.* employed a localized image metric to approximate nonrigid motion correction as multiple translational corrections (16,18). In their approach, so-called butterfly navigators were added to a 3D Cartesian acquisition providing motion estimates for each spatial direction and the signals from individual receiver coil channels were used to constrain the possible reconstructed motion states. In our work, the use of the self-navigation signal allows us to estimate motion without significantly modifying the imaging sequence. Additionally, the method proposed in this work for joint optimization of the image metric and motion map smoothness is similar to improvements discussed but not implement in previous work (16,17).

A key feature of the proposed fNAV approach is the ability to convert a unitless 1D signal into regional, millimetric 3D displacements. We can do this by leveraging the rich 3D isotropic information in our resulting images which lend themselves to quantitative evaluation by our chosen image metric. Our results suggest that this works very well in regions containing cardiac anatomy where the visibility of complex structures is sensitive to respiratory motion and consequently can be captured by the image metric. Conversely, regions in the image with less complex structures or relatively uniform signal are less sensitive to change and may be misinterpreted by the image metric, which may weaken our ability to estimate 3D motion in regions such as the lungs or chest wall, for example.

Alternative methods for measuring motion during the acquisition include separately acquired two-dimensional (2D) (34–37), and three-dimensional (3D) (20,38–43) image-based navigators. In particular, the works of Ingle *et al.* and Luo *et al.* have demonstrated that image navigators can be combined with autofocusing to reconstruct motion corrected data acquired using 3D spiral cones (17,20,41). While, these multidimensional approaches inherently provide more information about the underlying motion, they require interruption of the imaging sequence to acquire the navigators and typically do not have the required resolution for direct estimation of nonrigid motion. For ECG-triggered coronary imaging, the sequence is already interrupted and therefor a non-issue but recently proposed methods for "free-running" CMRA may benefit from the fNAV approach as they already periodically acquire SI readouts (21,44,45). 3D respiratory information can also be derived through nonrigid co-registration of respiratory resolved images (46,47) (48–51) but may be limited to inter-acquisition corrections in image space unless combined with a generalized matrix model (33,52,53) at the cost of increased computational complexity. For even further flexibility, our proposed approach could in principle be combined with a respiratory signal that is independent of the MRI sequence such as an external respiratory belt, or the recently proposed Pilot Tone Navigation System (54) which in turn facilitate motion correction using other types of sequences.

Our results clearly demonstrate that fNAV provides improved image quality relative to previously described methods for 3D radial CMRA and the two example comparisons to coronary angiography show the potential for evaluation of coronary disease using fNAV. Consequently, further studies are needed to investigate the impact of these improvements on clinically relevant parameters such as identification of stenoses and anomalous coronary arteries, as well as emerging techniques such as 3D modeling of the coronary anatomy for subsequent CMRA-based fractional flow reserve estimations (55). Furthermore, while the focus of this work was on coronary imaging, the employed 3D radial sequence provides whole-heart coverage with high isotropic spatial resolution and therefore the proposed fNAV approach may also be useful for evaluating other abnormalities in the cardiac anatomy such as congenital heart defects.

## Limitations and Future Directions

Validation of the fNAV framework including the identification of optimum tuning parameters was largely performed using a numerical simulation developed for this work. The goal of the numerical simulation was to synthesize data with sequence parameters that match *in vivo* acquisitions and simulate a realistic range of respiratory motion and heart rate variability. Still, limitations of this simulation framework include the appearance of the coronary arteries which are relatively large and well defined when compared to *in vivo* MR data sets because the XCAT models are derived from computed tomography data. Similarly, only one model of "normal" cardiac anatomy was used which does not reflect the variation we see in both healthy volunteers and patients. Finally, the MR physics simulated in this model consisted of a simplified equation for bSSFP contrast and complex coil sensitivities, neglecting the additional sources of artifact as well as effects of the T2-prep, saturation-slab, and fat saturation pulses that are used in the *in vivo* sequence.

Overall, the simulation provided a useful means of developing the fNAV framework, assessing the effects of motion and characterizing errors in the fNAV reconstructions, and the simulated results were generally corroborated by the large cohort of *in vivo* data that were analyzed. Nevertheless, improvements to the simulation framework or the addition of another ground truth measure may yield further improved tuning of the parameters for the fNAV framework. For example, an inherent trade-off exists when selecting the width (b) of the localized image metric in eq. 4. If the value of b is too small, noise-like artifact can appear in the derived displacement fields and subsequent images while a large value of b may fail to capture nonrigid deformation (16,17,19). Additionally, the number of motion states considered for the final fNAV reconstruction has a large impact on the ability to assess non-linearities in the displacement fields. This is particularly evident in the displacement fields shown in Fig. 2a where motion of the chest wall, for example, is not well represented by the fNAV reconstructions of simulated data.

Analysis of image quality across all four reconstruction methods was performed quantitatively using measurements of vessel sharpness and length and qualitatively via the best selected image for a given subject as identified by an expert reviewer. The rationale to choose only the best image was based on the need to evaluate 571 *in vivo* data sets. However, in subjects with minimal motion it may be difficult to identify a single best image and therefore subjective image grading may have been more appropriate. Nevertheless, the current qualitative results agree with the overall narrative provided by the quantitative results that show fNAV provides the overall best image quality.

Finally, in the current model, we assume that different regions of the image are affected by different respiratory motion amplitudes, but we do not consider the phase. As such, hysteresis effects may contribute to errors in our estimated displacement fields and degrade our image quality. This may be addressed by including additional parameters in our fNAV model (i.e. respiratory phases coefficients for each spatial direction) but will potentially increase the computation time if more motion states need to be considered. Regardless, the inclusion of additional spatial information such as self-navigation signals from individual receiver coils may help constrain the potential motion states (16,56).

## Conclusion

Focused navigation is a promising technique for improving the quality of self-navigated free-breathing 3D radial whole-heart CMRA. This novel approach to respiratory self-navigation can derive 3D nonrigid motion estimations from an acquired 1D signal yielding statistically significant improvement in image sharpness relative to 1D translational correction as well as XD-GRASP reconstructions. Using this approach motion corrected images can be reconstructed in ~5-10 mins compared to ~20-30 minutes using XD-GRASP, potentially facilitating integration in a clinical environment. Further study of the diagnostic impact of this technique is therefore warranted to evaluate its full clinical utility.

# Tables

**Table 1:** Simulated and *in vivo* acquisition parameters.

| Parameter | Simulations | Volunteers | Patients |
|---|---|---|---|
| **Field-of-view (mm$^3$)** | 220 | 200 - 210 | 185 – 250 |
| **Voxel size (mm$^3$)** | 1 | 0.98 - 1.09 | 0.88 – 1.15 |
| **Excitation angle (°)** | 100 | 90 - 115 | 60 - 115 |
| **Repetition time (ms)** | 3.14 | 2.97 - 3.33 | 2.5 – 5.27 |
| **Echo time (ms)** | 1.6 | 1.57 - 1.76 | 1.56 – 1.79 |
| **Radial interleaves** | 560 | 191 - 492 | 199 - 2351 |
| **Profiles per interleave** | 26 | 25 - 69 | 8 - 60 |
| **Acquisition time (s)** | 475-843 | 219 – 363 | 167 – 917 |

# Figures

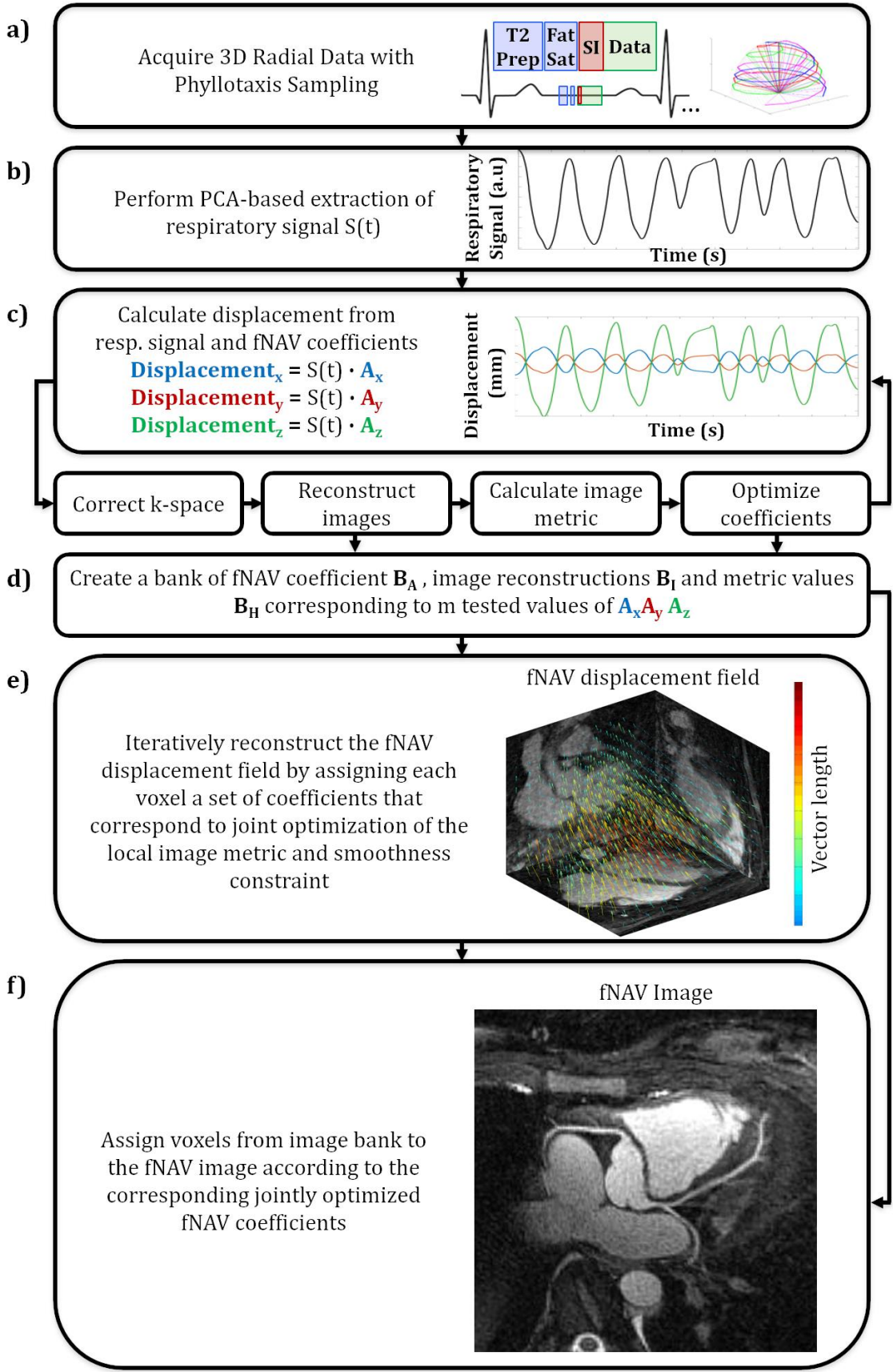

**Fig. 1** Schematic overview of the focused navigation framework

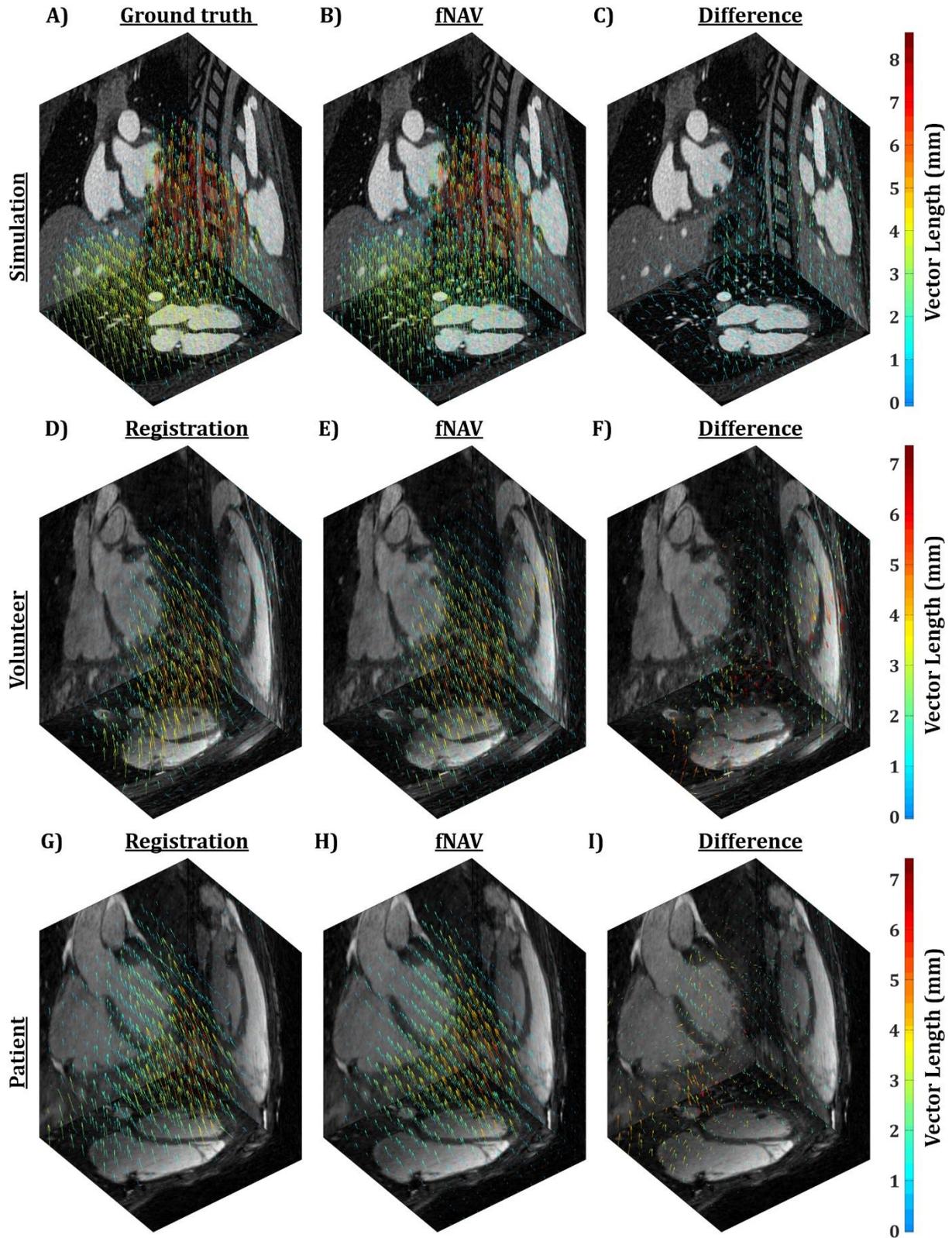

**Fig. 2 Vector representation of respiratory motion**. Displacement fields derived from ground truth (A) and fNAV reconstructions of simulated data (B) demonstrate the nonrigid behavior of respiratory motion during end-inspiration, and excellent visual agreement is observed between fNAV and ground truth as shown by the difference image (C). Similarly, displacement fields derived from registering the frames of XD-GRASP reconstructions of a representative healthy volunteer (D) and patient (G) are comparable to those derived using fNAV (E & H) as demonstrated by their corresponding difference images (F & I).

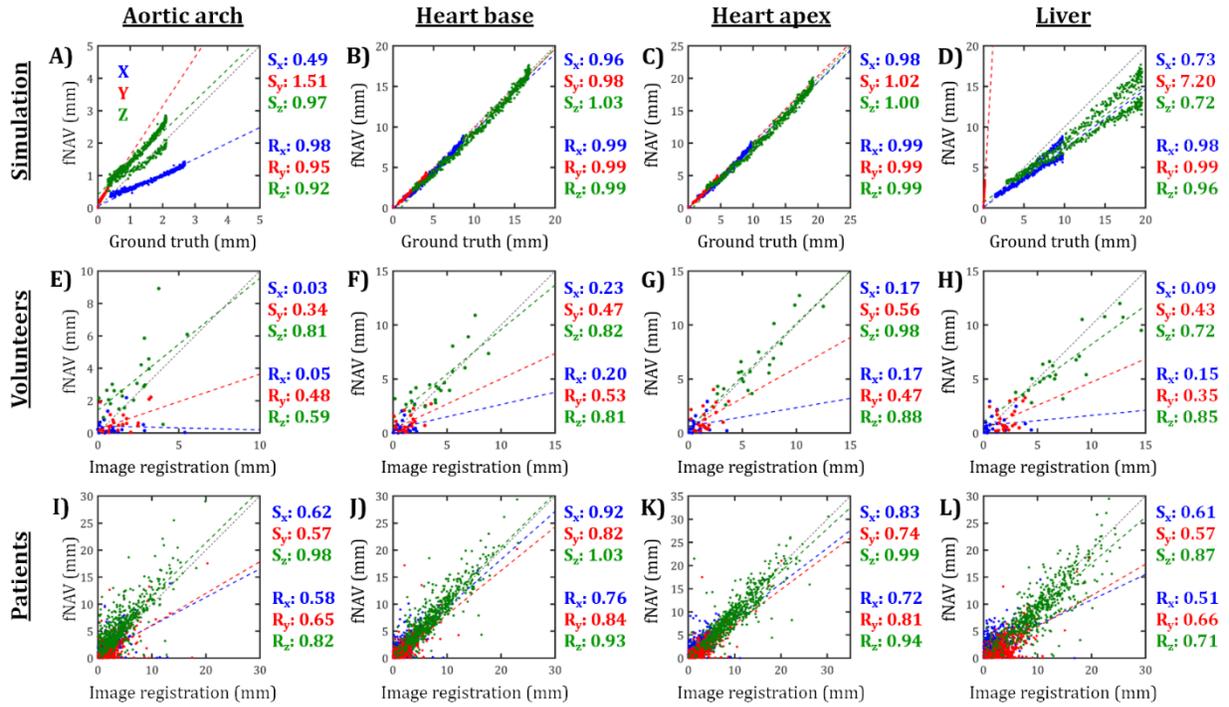

**Fig. 3 Quantitative evaluation of respiratory motion estimation.** The mean x, y, and z components of respiratory motion estimated by fNAV measured in four regions of interest are shown for all 500 simulated data sets (A-D), all 22 healthy volunteer data sets (E-H). and all 549 patient data sets (I-L). Motion estimates derived from fNAV are compared to ground truth, and estimates derived from image registration for simulated and *in vivo* data, respectively. The results of a linear fit for each component is denoted by dash lines with the corresponding slopes and Pearson correlation coefficients given by [$S_x$, $S_y$, $S_z$] and [$R_x$, $R_y$, $R_z$] respectively.

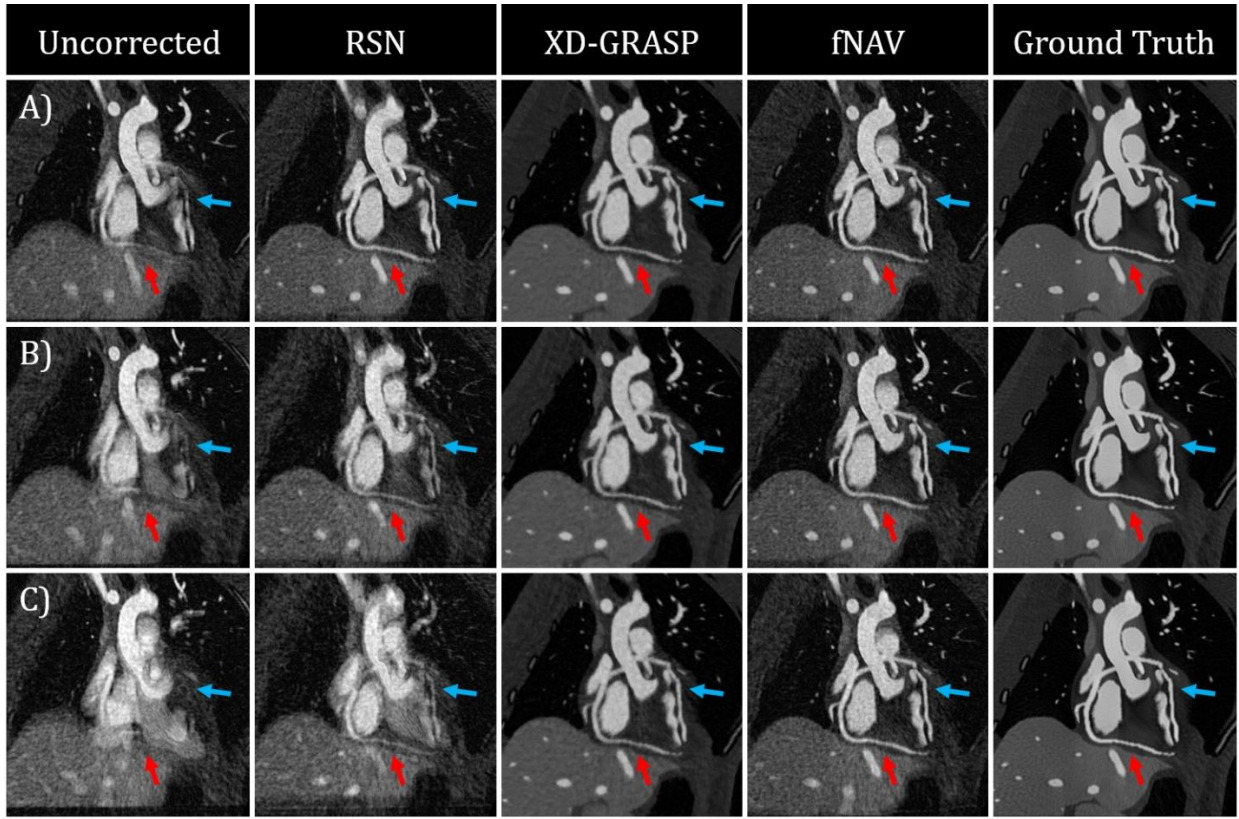

**Fig. 4 Curved reformats of simulated data reconstructions at increasing levels of simulated motion.** A) Respiratory motion in only one dimension (maximum amplitudes LR = 0 mm, AP = 0 mm, SI = 10 mm). B) Respiratory motion in two dimensions (maximum amplitudes LR = 0 mm, AP = 5 mm, SI = 10 mm). C) Respiratory motion in thee dimensions (maximum amplitudes LR = 5 mm, AP = 10 mm, SI = 20 mm). Arrows denote the right (red) and left anterior descending (blue) coronary arteries.

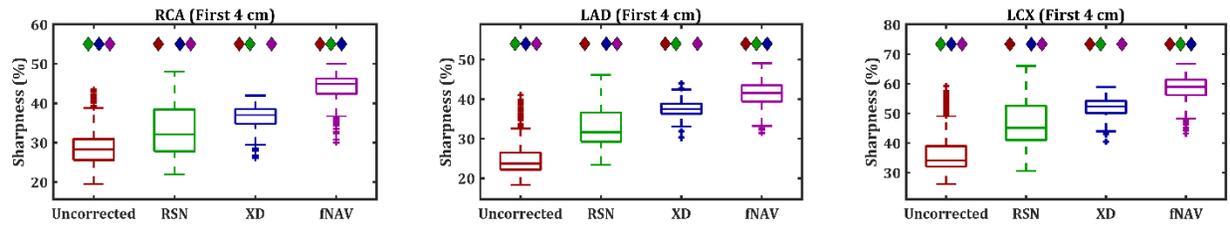

**Fig. 5 Quantitative evaluation of vessel sharpness from simulated data.** Percent vessel sharpness measured across the full length of the RCA (left), LAD (middle), and LCX (right) in all 500 simulated data sets is shown. Colored diamonds denote statistically significant differences between the four reconstruction methods (p < 0.01).

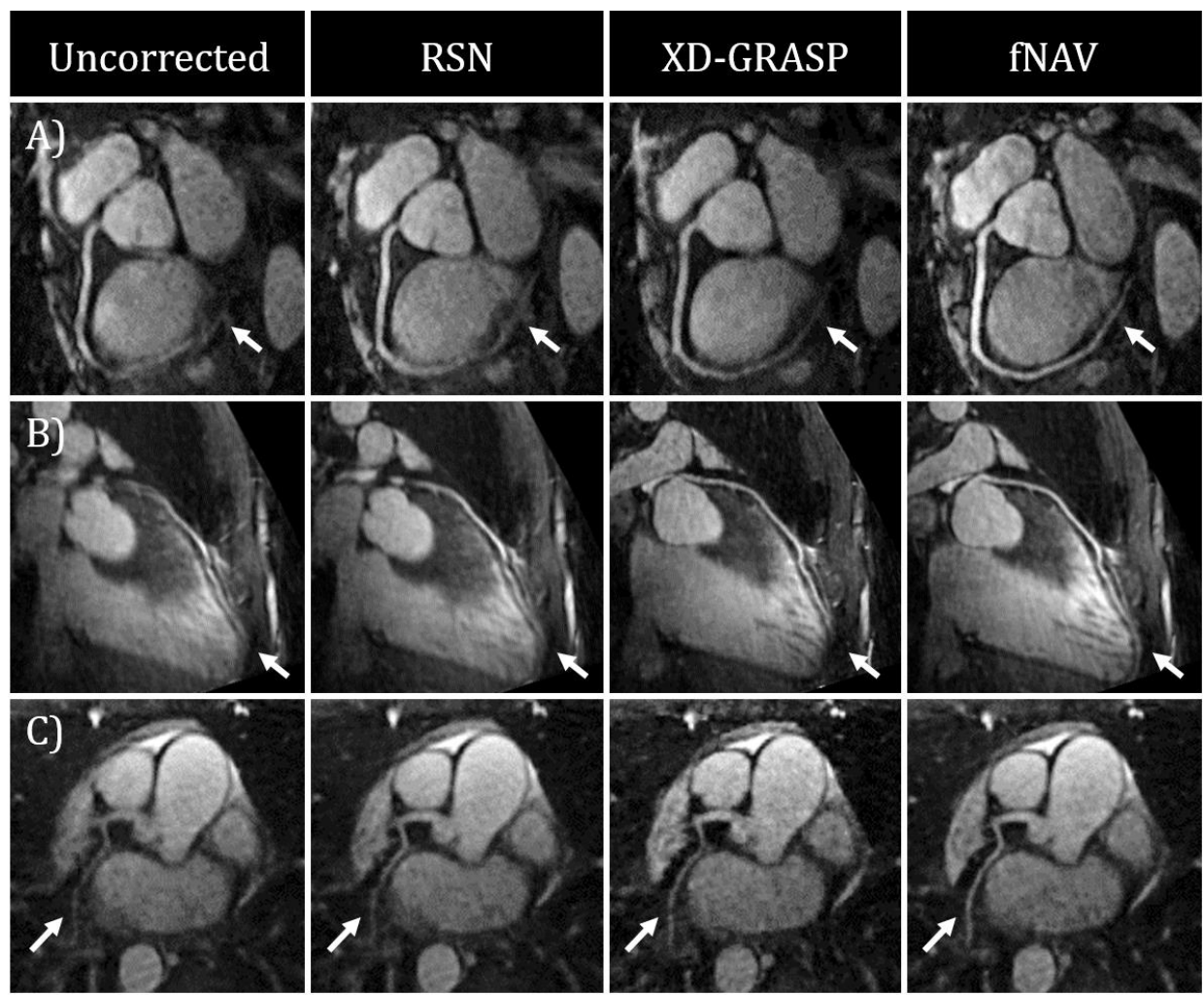

**Fig. 6 Curved reformats of healthy volunteer data.** The RCA (A), LAD (B), and LCX (C) are shown from a three representative volunteers with white arrows denoting regions of the vessels where image quality varies between the four reconstruction methods.

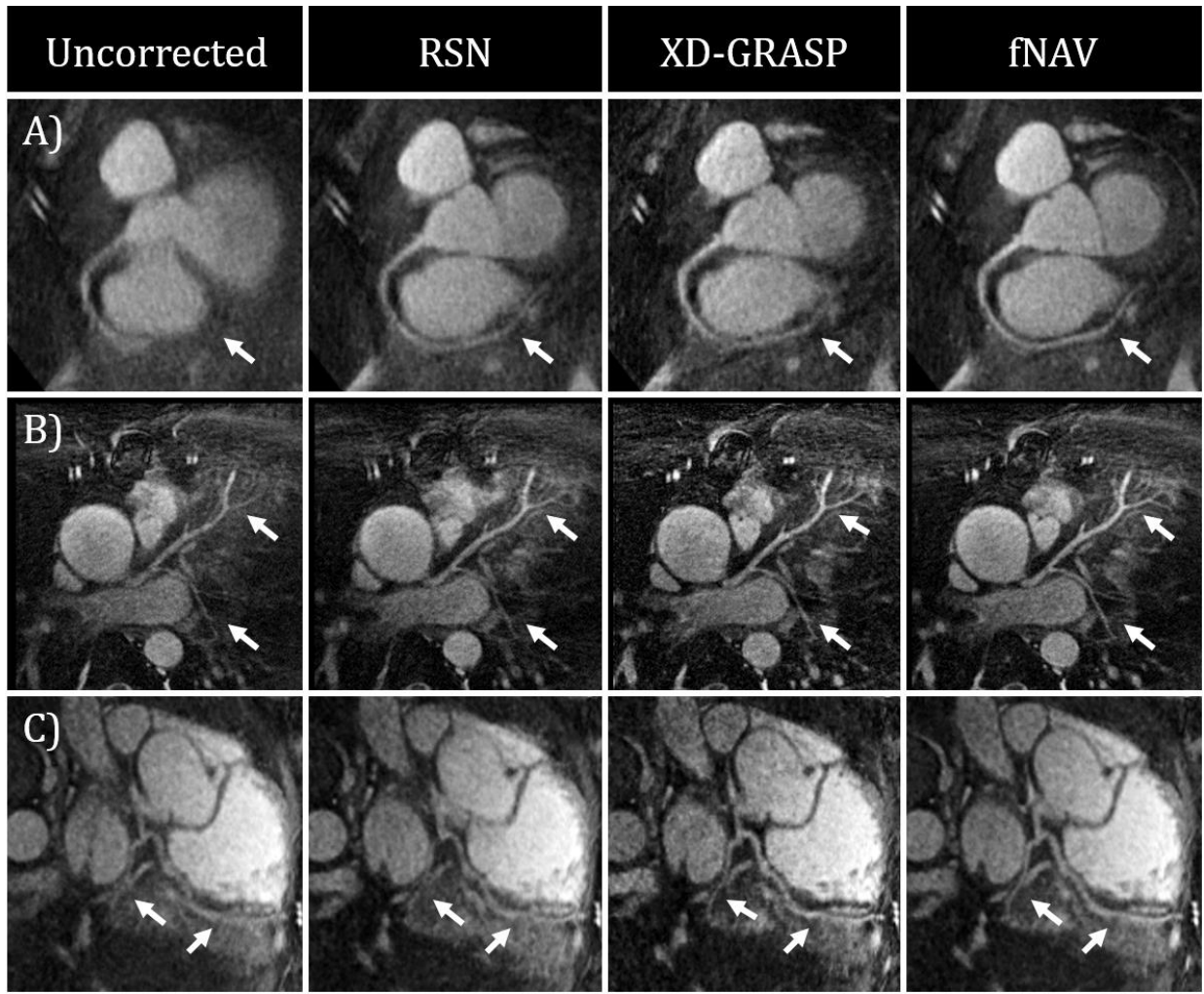

**Fig. 7 Curved reformats of patient data.** The RCA (A) and LAD+LCX (B & C) are shown from three representative patients with white arrows denoting regions of the vessels where image quality varies between the four reconstruction methods.

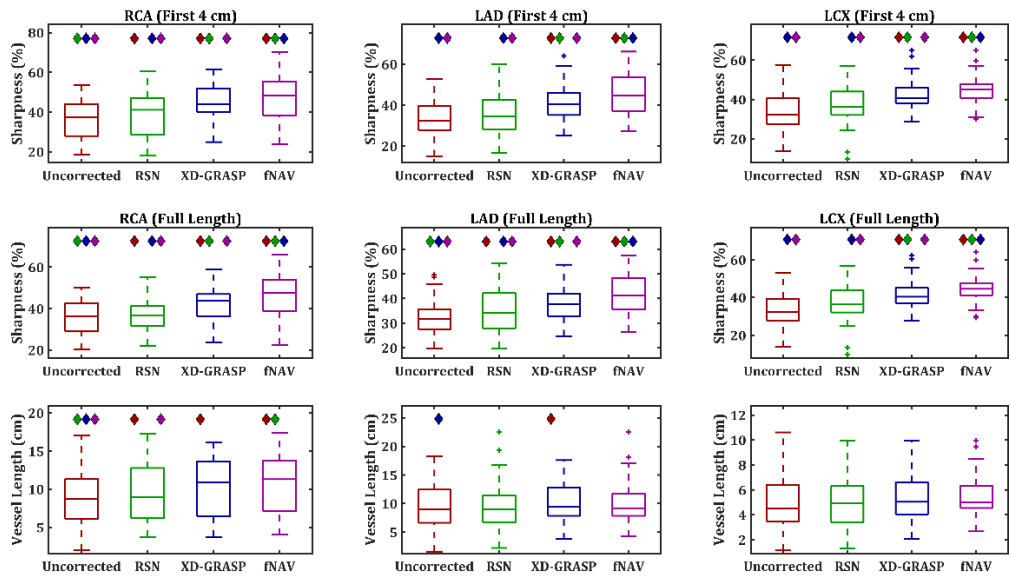

**Fig. 8 Quantitative evaluation of vessel sharpness and length from all volunteers (n = 22) and a subset of patients (n = 20).** Percent vessel sharpness measured across the first 4 cm (top row) and full length (middle row) of the RCA (left), LAD (middle), and LCX (right) as well as vessel length (bottom row) is shown. Colored diamonds denote statistically significant differences between the four reconstruction methods.

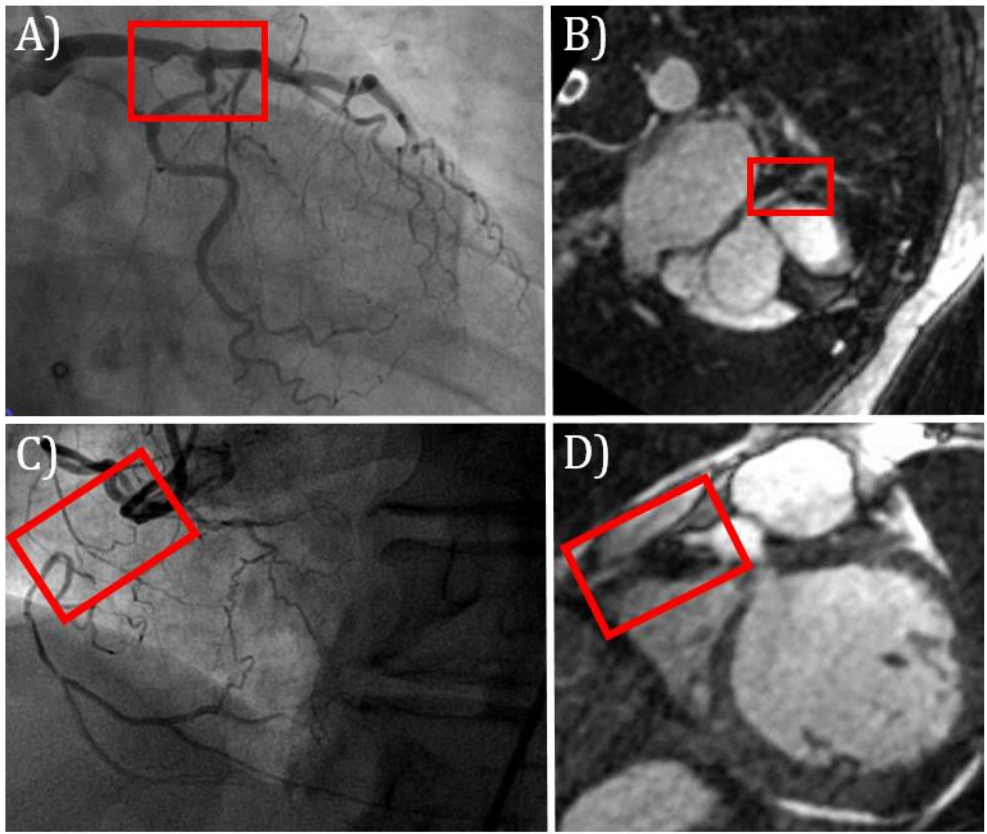

**Fig. 9 Visualization of coronary artery disease using fNAV.** Critical stenosis of the left main shown by coronary angiography (A) and corresponding fNAV image (B). Chronic total occlusion of the proximal RCA shown by coronary angiography (C) and corresponding fNAV image (D).